\documentclass[a4paper,fleqn,usenatbib]{mnras}

\usepackage[T1]{fontenc}
\usepackage{ae,aecompl}
\usepackage{xcolor}

\newcommand{\lsim}{\mathrel{\rlap{\raise -.3ex\hbox{${\scriptstyle\sim}$}}%
		  \raise .6ex\hbox{${\scriptstyle <}$}}}%
\newcommand{\gsim}{\mathrel{\rlap{\raise -.3ex\hbox{${\scriptstyle\sim}$}}%
		  \raise .6ex\hbox{${\scriptstyle >}$}}}%

\usepackage{graphicx}	
\usepackage{amsmath}	
\usepackage{amssymb}	

\title[Nighttime photometry of the anthropogenic skyglow]{Nighttime monitoring of the aerosol content 
of the lower atmosphere by differential photometry of the anthropogenic skyglow}

\author[M. Kocifaj et al.]{
Miroslav Kocifaj,$^{1,2}$\thanks{E-mail: kocifaj@savba.sk}
Salvador Bar\'{a},$^{3}$
\\
$^{1}$Faculty of Mathematics, Physics, and Informatics, Comenius University, Mlynsk\'{a} dolina, 842 48 Bratislava, Slovakia\\
$^{2}$ICA, Slovak Academy of Sciences, D\'{u}bravsk\'{a} cesta 9, 845 03 Bratislava, Slovakia\\
$^{3}$Departamento de F\'{\i}sica Aplicada, Universidade de Santiago de Compostela, 15782 Santiago de Compostela, Galicia, Spain
}

\date{Accepted XXX. Received YYY; in original form ZZZ}

\pubyear{2020}

\begin{document}
\label{firstpage}
\pagerange{\pageref{firstpage}--\pageref{lastpage}}
\maketitle

\begin{abstract}
Nighttime monitoring of the aerosol content of the lower atmosphere is a challenging 
task, because appropriate reference natural light sources are lacking. Here we show that the anthropogenic night sky brightness due to city lights can be successfully used for estimating the aerosol optical depth of 
arbitrarily thick atmospheric layers. This method requires measuring 
the zenith night sky brightness with two detectors located 
at the limiting layer altitudes. Combined with an estimate of the overall 
atmospheric optical depth (available from ground-based measurements or specific 
satellite products), the ratio of these radiances provides a direct estimate of the 
differential aerosol optical depth of the air column between these two altitudes. These 
measurements can be made with single-channel low-cost radiance detectors widely 
used by the light pollution research community.

\end{abstract}




\section{Introduction}

Monitoring the aerosol content of the lower atmosphere is relevant 
for characterizing the atmospheric boundary layer dynamics. Nighttime aerosol 
observations, however, are significantly more demanding than their daytime 
counterparts, specially if passive techniques (i.e. those based on light not 
emitted by the observer) are to be used. Whilst during daytime the Sun provides a 
reliable and radiometrically well characterized light source allowing for continuous aerosol 
monitoring \citep[]{Aeronet2020}, the night lacks such a natural lighting standard with the required attributes.

City lights offer an alternative source of photons to sample the 
nocturnal atmosphere. Artificial light emitted from urban and rural nuclei, roadway infrastructure, and diverse industrial, agricultural or extractive facilities, is a pervasive feature of the modern world. A 
non-negligible amount propagates toward the upper hemisphere, either directly 
or after being reflected on different natural and artificial surfaces. 
A fraction of this light leaves the upper atmosphere and can be analysed 
by spaceborne instruments. Other fraction propagates downward, being perceived by ground-based observers as light pollution in form of 
increased sky brightness (anthropogenic skyglow). Both light beams interact with 
the atmospheric constituents through spatially distributed processes of absorption and 
scattering, and contain information about their composition and volume distribution, 
predominately aerosols \citep[]{SciezorCzaplicka2020}. 

Remote sensing of city lights from low-Earth orbit has been recently proposed as a useful tool for atmospheric 
research. Direct line-of-sight radiometry of urban lights 
allows to determine, among other parameters, the aerosol optical depth ($AOD$) 
\citep[]{ChooJeong2016,JohnsonEtAl2013,McHardyEtAl2015,WangEtAl2016,ZhangEtAl2008,ZhangEtAl2019}. 
The halos of scattered light observed from space around ground light sources also 
provide interesting information. \citet[]{SanchezDeMiguelEtAl2020} have shown that their 
scattered upward radiance is correlated with the downward one, and hence with 
the light perceived by ground-based observers as anthropogenic skyglow, thus providing a 
feasible tool for wide-area light pollution monitoring. Analysing these halos, 
\citet[]{KocifajBara2020} developed a method for estimating the number-size distribution of 
aerosol particles in the nocturnal air column from the angular distribution of the upward 
scattered light.

We describe here a new application, namely a practical approach for determining 
the difference in aerosol optical depth ($\Delta AOD$) between two altitudes in the 
nocturnal air column, based on differential photometry of the artificial skyglow. 
$\Delta AOD$ could be derived from ground $AOD$ if the aerosol 
scale height for the exponential atmosphere were known. However, the instantaneous aerosol scale 
height is generally unavailable for a specific site, and using mean values can 
result in considerable errors. The method here described allows estimating 
$\Delta AOD$ from measurements of the zenith sky brightness made at two different altitudes, in the vicinity of the urban centres, and having a good estimate of the total $AOD$. The ratio between these
radiances provides a good estimate of the fraction of the total $AOD$ that is due to the 
aerosols in the layer between the two measuring heights. ($\Delta AOD$) is important for characterizing the nocturnal contamination of the lower 
atmosphere, which is typically unknown because many detectors (like those at AERONET 
stations) are designed to monitor columnar aerosol optical properties 
\citep[]{KimEtAl2007} and primarily operate in daytime. This monitoring network has also 
taken advantage of moonlight \citep[]{BerkoffEtAl2011}, but these data are 
available at a few stations and are mostly limited to the nights near full moon. Since the data coverage is extremely uneven and sparse (we notice there is 
only one AERONET station in Slovakia), the use of this product is restricted 
to relatively few sites. The measurements required by our approach can be made with relatively low-cost instruments widely used by the light pollution research 
community \citep[]{HanelEtAl2018}. We expect this method can contribute to increase the amount and the quality of aerosol data in the nocturnal atmosphere.

\section{Modeling the dependence of the zenith sky brighness with altitude}
\label{sec:model}

The artificial night sky radiance is due to the backscattering of the city light emissions in the Earth 
atmosphere. 
In a random multiple scattering atmosphere the NSB can be obtained by solving the radiative 
transfer equation with adequate boundary conditions. However, for optically thin 
media with total optical depth not exceeding unity, the scattered light field may be 
advantageously decomposed into scattering modes, a procedure known as the {\it method 
of successive orders}. This approach is particularly useful when the first scattering 
mode dominates the remaing ones, as e.g when the light beams travel short optical paths \citep[]{Kocifaj2018}. This is the usual case when the NSB is observed close to an artificial light source. For example, 
in mountain regions under clear skies and $AOD$ below 0.3, the expected second-scattering 
radiance at the zenith ranges from 2\% to 9\% of the radiance of the single scattering 
component (based on own computations, not shown here). In these cases the NSB 
can be computed much faster using a single-scattering approximation than by superimposing many high-order modes (ibidem and in \citet[]{Aube2015}). 
We have shown in \citep[]{Kocifaj2018} that the first and higher scattering radiance components vary strongly with the atmospheric optical depth. 

Let us recall that the aerosol optical depth ($AOD$) and the Rayleigh optical depth ($ROD$) 
are defined for a vertical beam path from a given altitude $h$ to the top of the atmosphere. $AOD$ or $ROD$ are  
the natural logarithms of the incident-to-transmitted radiant power for an aerosol or a 
molecular atmosphere, separately. The scattering phase function, in turn, characterizes 
the angular distribution of the radiance scattered by a small atmospheric 
volume. This is a wavelength-dependent function with a large anisotropy in 
turbid media with high concentration of aerosols of sizes comparable to (or larger than) the wavelength of light 
\citep[]{Mishchenko2009}. The particles whose sizes and optical properties tend to 
differ from those of Rayleigh scatterers are often called Mie scatterers (although the 
conventional Mie theory is exclusively applicable to homogeneous spherical particles). 
Their shapes can be arbitrary and their sizes are too large for being 
approximated by a single pointlike dipole source. To compute their far-field scattered 
radiance each particle shall be replaced 
by large array of point dipoles. The superposition of all 
scattered waves, with different relative phases, tends to produce a 
complex scattering pattern. Hence the scattering phase function is expected to vary 
greatly with the microphysical properties of aerosols, specifically their sizes, 
shapes, and compositions, but also with the specific way in which the individual 
particles are distributed within the scattering volume. 

Not all the above mentioned parameters have the same role in shaping the patterns of diffuse light at night. Some of them 
are particularly significant and should be determined with care, specially when they vary 
quickly in time. The spectral ground reflectance may be very different 
from one city to another and shows relevant seasonal trends (permanent 
snow cover, leaf litter...), but, excepting for transient episodes of rain or snow 
coating, it may be expected to be relatively constant at the city scale across any 
typical single night.
The aerosol system, however, plays a key role. It is highly unstable and sensitive to 
meteorological conditions at the observing site, and often undergoes abrupt changes in 
short characteristic times, as e.g. with wind direction or passage of atmosperic fronts. The aerosol microphysics also changes 
with air humidity, local production, transport and chemical transformation of particles, and 
hence the scattering and absorption properties of aerosols tend to vary from day to day 
and even within a single day in any particular site. Moonlight may also influence the 
measurements significantly \citep[]{PuschnigEtAl2020,CavazzaniEtAl2020a,CavazzaniEtAl2020b}
by reshaping the radiance patterns, therefore it is highly preferred to make the 
observations during moonless nights or when the Moon is low.

For many applications we are interested in monitoring the NSB changes in sites whose parameters are reasonably constant or have a well known and 
deterministic temporal variation. This applies notably to the emissions of the city lights: 
although the spatial distribution, composition, orientations and angular emission functions 
of individual light sources are generally unknown for a city, these factors are fairly 
constant or vary predictably across a single night (see, e.g. 
\citet[]{DoblerEtAl2015,Meier2018,BaraEtAl2019a}). Among all site-specific parameters, we 
have found that the optical depth is the most relevant one when it comes to determining 
the artificial zenith radiance. We have performed a number of numerical experiments 
to select the variables most influential on the NSB changes recorded 
at two different altitudes. In spite of potential discrepancies between 
the theoretically assumed atmospheric profiles and the actual ones, we have 
found that the ratio of zenith radiances measured by a ground-based station and other 
located at an altitude $h$ above ground level is basically driven by the optical thickness 
of the lower atmospheric layer comprised between the two stations, $\Delta \tau_{0-h}$, and by the 
optical depth of the atmosphere above the highest one, $\tau_{h}$. Henceforth we will use 
the term $\Delta OD=\Delta AOD+\Delta ROD$, the differential optical depth, as an equivalent 
notation for $\Delta \tau_{0-h}$.

\begin{figure}
	\begin{center}
	\includegraphics[width=0.85\columnwidth]{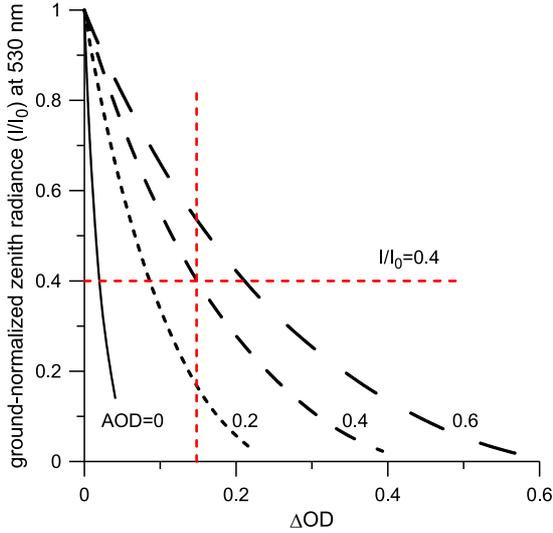}
	\end{center}
    \caption{Zenith radiance ($I$) of the night sky at altitude $h$, relative 
	to its value at ground-level ($I_{0}$). The ratio of $I/I_{0}$ is shown as a function 
	of $\Delta OD$, the optical thickness of the lower atmospheric layer between ground 
	level and the elevated station. Computations are made for different values of the columnar 
	aerosol optical depths ($AOD$s). The point at which the vertical and horizontal red (dashed) 
	lines intersect shows how $\Delta OD$ can be inferred from the measurements $I/I_{0}$=0.4 and $AOD$=0.4. The latter is available as a satellite-derived product, 
	while $I/I_{0}$ is obtained as the ratio of the zenith radiance at the two altitudes. 
	The line for $AOD$=0 corresponds to pure molecular scattering.}
	\label{fig:Fig1}
\end{figure}

The ground-normalized zenith radiance, defined as the ratio of the zenith radiance at altitude ${h}$ ($I$) to its ground-level counterpart ($I_{0}$), is shown in 
Fig.~\ref{fig:Fig1} as a function of the optical depth of the lower atmospheric layer, 
$\Delta OD$. The calculations were made for a narrow spectral band centered at 
530 nm (the average wavelength of the device spectral responsivity) using the 
theoretical model published in \citep[]{Kocifaj2018}. The results are given for a set of 
typical values of the ground level $AOD$, the columnar aerosol optical depth, a commonly available satellite product 
\citep[]{ShahzadEtAl2018,HasekampEtAl2019,ChongEtAl2019}. 
The numerical calculations have been 
made including five scattering orders, using the expressions (1,6,14-15) of the theoretical
model described in \citep[]{Kocifaj2018}. This number of scattering orders allow us obtaining 
reliable results even for atmospheres with high aerosol content, and hence large $AOD$ values. The results are shown as a function of $\Delta OD$ for $AOD$ ranging from 0 to 0.6.
It can be seen in Fig.~\ref{fig:Fig1} that the differential optical depth $\Delta OD$ between 
both stations can be univocally determined for any $AOD$ value 
from the measured ground-normalized zenith radiance $I/I_{0}$. For instance, if both 
$I/I_{0}$ and $AOD$ are 0.4 then $\Delta OD \approx$0.15, as indicated in the Figure. 

\section{An approximate formula for the ground-normalized zenith brightness}
\label{sec:theory}

\begin{figure}
	\begin{center}
	\includegraphics[width=1\columnwidth]{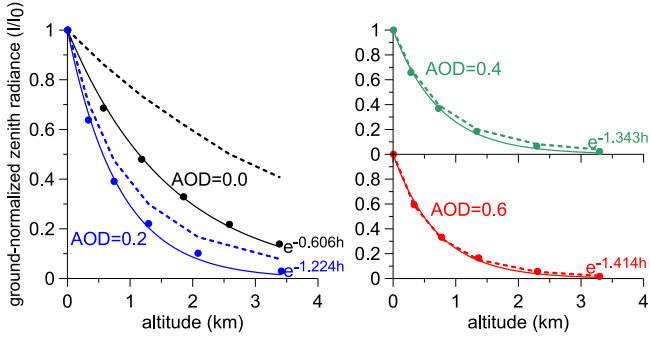}
	\end{center}
    \caption{Ground-normalized zenith radiance $I/I_{0}$ for a set of  $AOD$ as determined 
	from numerical integration of the light propagation equations (dots); the approximate formula 
	Eq.~\ref{eq:approximation} (dashed lines); and the empirical formula Eq.~\ref{eq:empirical} 
	(solid lines). The exponential expressions correspond to the best fits to the 
	numerically computed data.}
	\label{fig:Fig2}
\end{figure}

Air molecules and particles much smaller than the wavelength of light scatter 
in the Rayleigh regime, with a scattering efficiency inversely proportional 
to the fourth power of the wavelength, and a forward scattering lobe identical to the backward 
one. The scattering phase function is anisotropic for most kind of aerosols, and their wavelength dependence is smaller than for Rayleigh scatterrers. 
Therefore, the largest differences in the underlying physics can be expected to occur when
transitioning from an aerosol-free atmosphere to slightly turbid one. Assuming that the turbid 
atmosphere is lit from below by a huge amount of point light sources we have found, after 
lenghty mathematical derivations, that the ground-normalized zenith radiance near a city can 
be approximated by the formula:
\begin{eqnarray}\label{eq:approximation}
    \frac{I}{I_{0}} &=& \frac{\tau_{h}}{\tau_{0}} e^{\tau_{h}-\tau_{0}}~
	\frac{e^{\tau_{h}}-e^{-\tau_{h}}}{e^{\tau_{0}}-e^{-\tau_{0}}},
\end{eqnarray}
where $\tau_{h}$ is the atmospheric optical depth above the elevated station, and $\tau_{0}=AOD+ROD$ 
is the corresponding value at ground level. The optical thickness of the lower atmospheric layer 
is $\tau_{0}-\tau_{h}=\Delta \tau_{0-h}=\Delta OD=\Delta AOD+\Delta ROD$. Eq.~\ref{eq:approximation}
shows a progressively better coincidence with the exact theoretical predictions when the 
$AOD$ values increase. This is shown in Fig.~\ref{fig:Fig2}, where the dashed lines correspond 
to Eq.~\ref{eq:approximation}, while the dots are the results of the numerical integration of the 
exact equations governing the artificial night sky radiance. The solid lines in this Figure correspond 
to the exponential functions that best fit the exact computational results (shown as dots as indicated 
above). Furthermore, assuming that the average $ROD$ at the effective instrumental wavelengths decreases 
exponentially with the altitude, $ROD \propto 0.118~e^{-h/8}$ ($h$ is in km), we found the following 
approximate expression for the ground-normalized zenith radiance:
\begin{eqnarray}\label{eq:empirical}
    \frac{I}{I_{0}} &\cong& e^{ah} \tau_{0}^{bh},
\end{eqnarray}
where $a \cong -1.64$ $km^{-1}$ and $b \cong -0.46$ $km^{-1}$. In the formula for
$ROD$ the coefficient 0.118 is the theoretical $ROD$ at sea level, while the 
value 8 in the exponent is the altitude (km) up to which a homogeneous molecular atmosphere 
would extend. Fig.~\ref{fig:Fig2} also shows that the ratio $I/I_{0}$ modelled from Eq.~
\ref{eq:approximation} tends to deviate significantly from the numerical exact values when 
the $AOD$ is approaching zero (i.e. approaching an aerosol-free, purely molecular atmosphere). In an 
aerosol-free atmosphere the average discrepancy between the numerical and the approximate values is 
25\%, reducing to only 7\% for $AOD$=0.2, and just 2\% for $AOD$=0.4. Let us finally note 
that $I/I_{0}$, as per Eq.~\ref{eq:approximation}, decreases proportionally 
to the square of $\tau_{h}$ as $h$ increases, since $I/I_{0}$ asymptotically approaches 
$2 \frac{\tau_{h}^{2}}{\tau_{0}} \left ( \frac{1}{e^{2\tau_{0}}-1}\right )$. The 
formula is obtained from Eq.~\ref{eq:approximation} by applying Taylor-series expansion to the 
exponential functions of $\tau_{h}$, assuming $\tau_{h}$ is a small value at elevated altitudes. 
In turn, $I/I_{0} \rightarrow 1$ when $\tau_{h} \rightarrow \tau_{0}$, i.e, when the altitude $h$ of 
the elevated station approaches that of the base one.

\section{Experimental results from a field measurement campaign in a mountain region}
\label{sec:experiment}

The required $I/I_{0}$ normalized radiance can be measured with relatively simple 
broadband single-channel radiance detectors, as e.g. the widely used Sky Quality Meter (SQM) 
of Unihedron (Grimsby, ON, Canada), and the Telescope Encoder and Sky Sensor (TESS-W) developed by the 
European Union's Stars4All project \citep[]{PravettoniEtAl2016,ZamoranoEtAl2016,HanelEtAl2018,BaraEtAl2019b}. The linear distance between the base station and the 
elevated one should be at least a few hundred meters, to achieve the instrumental detection limit for a 
measurable NSB change. 

\begin{figure}
	\begin{center}
	\includegraphics[width=1\columnwidth]{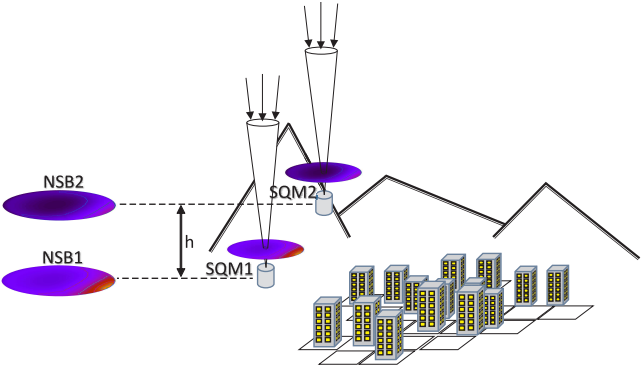}
	\end{center}
    \caption{Configuration of the field experiment in a mountain region. The vertical separation 
	($h$) between the base (SQM1) and the elevated station (SQM2) is chosen to be a few hundred 
	meters. The city emits light upwards, while the detectors (SQM1 
	and SQM2) measure the backscattered light from the zenith region, within a Gaussian field-of-view 
	with full-width at half-maximum (FWHM) 20$^{\circ}$. NSB1 and NSB2 are the hemispherical night 
	sky brightness distributions at both stations, projected onto a horizontal plane. The zenith is 
	located at the center of each projection.}
	\label{fig:Fig3}
\end{figure}

A field experiment has been carried out in the High Tatra region (Slovakia) during late autumn 2019 
(from Nov. 14 to Nov. 15) with snow cover absent. Since it was in the full moon
period we limited our experiment to the first hour of astronomical night (from 6:00 PM to 6:50 PM). 
The Sun was 18$^{\circ}$ to 26$^{\circ}$ below the horizon, while the Moon was between 2$^{\circ}$ 
and 10$^{\circ}$ above the horizon, however, physically out-of-sight because of the presence of 
clouds near the horizon and obstacle blocking at low elevation angles. The cloud-free region spanned 
a wide area of the sky, with stable clarity around zenith. The base station was 
close to Tatransk\'{a} Lomnica (850 m a.s.l.), a small town connected by cableway with 
the elevated station at Skalnat\'{e} pleso (1780 m a.s.l). Skalnat\'{e} pleso is the midpoint 
between the base station and Lomnick\'{y} peak. Located almost 900 m above the 
base station (see Fig.~\ref{fig:Fig3}) and small plateau around the astronomical observatory, 
Skalnat\'{e} pleso is an ideal platform to make the measurements suggested in this paper. 
The SQM1 sensor was installed in an open space aside the city, but close to the cableway (likewise SQM2).
The measured ground-normalized zenith radiance was $I/I_{0}=0.22 \pm 0.05$. The absolute maximum 
uncertainty of 0.05 was derived from the SQM error which is $\pm$ 10\% (as reported by manufacturer), 
while the value is doubled when determining the ratio of two radiances taken under different 
conditions by two different SQM devices (20\% of 0.22 is 0.044; we rounded it up to 0.05). 
The $AOD$ was obtained from the AERONET station located in Poprad-G\'{a}novce (49.03500$^{\circ}$ 
North, 20.32200$^{\circ}$ East), approximately 15 kilometres apart. This is advantageous since no satellite data were needed. The horizontal distance between the base 
station (SQM1: 49.166150$^{\circ}$N, 20.268868$^{\circ}$E) and the elevated station (SQM2: 
49.188333$^{\circ}$N, 20.232778$^{\circ}$E) is only 4 km, which make us believe that the aerosol changes 
are mostly due to vertical stratification and not to the geographical distance. 
Normally two distant sites (tens of kilometres away) may differ in aerosol 
systems because the aerosol properties tend to decorrelate at such distances. This effect seems to be of minor relevance for the small area analysed 
here. The city of Poprad is really the only important light- and air-pollution source 
in the vicinity of the measuring site (see Tab.~\ref{tab:statistics}).

\begin{table}
	\centering
	\caption{Geographic coordinates and altitudes of the SQM, AERONET station 
	and the main cities/towns present. The distance to the SQM1 installed at the base station is
	also provided.}
	\label{tab:statistics}
	\begin{center}
	\begin{tabular}{l|r|r|r|r} 
	\hline\hline               
	~ & latitude & longitude & elevation & distance \\ 
	~ & (North) & (East) & (m) & (km) \\ \hline
	SQM1 &  49.166150 & 20.268868 & 903 & - \\ 
	SQM2 &  49.188333 & 20.232778 & 1780 & 4 \\
	AERONET &  49.035000 & 20.322000 & 706 & 15 \\
	T. Lomnica &  49.165300 & 20.278800 & 850 & 1 \\
	Poprad &  49.059444 & 20.297500 & 672 & 12 \\
	Smokovce &  49.140974 & 20.221024 & 1010 & 5 \\
	\hline\hline
	\end{tabular}
	\end{center}
\end{table}

The measured value $I/I_{0}$=0.22 in combination with the AERONET record 
$AOD \approx$ 0.28 provides the final estimate $\Delta OD \approx 0.18 \pm 0.06$ (compare to 
Fig.~\ref{fig:Fig1}). The uncertainty margin is derived from the amplitude of the diurnal 
fluctuation of $AOD$, strongly related to the near-surface pollutant mixing in the area.
Due to the short horizontal distance between SQM1 and the AERONET station we assume that the $AOD$ 
uncertainty is comparable to the amplitude of its diurnal fluctuation. Although AERONET 
measurements in Poprad-G\'{a}novce are given for the wavelength 500 nm, the conversion to 530 nm is simple through \AA ngstr\"{o}m exponent $\alpha$. The values of $AOD$ 
at 530 nm and 500 nm are almost identical since $\alpha$ was as low as 0.06 on Nov. 14, 2019.
The value of the $\Delta AOD \approx 0.17$ was inferred from 
$\Delta OD$, taking into account $\Delta ROD=0.118~(e^{-h_{1}/8}-e^{-h_{2}/8} )$ is 0.011 
for $h_{1}$=0.9 km and $h_{2}$=1.78 km. Although the SQM1 site could be classified as a high-elevation station, the $AOD$ values are not low because the mean 
altitude of the surrounding terrain is about 700 m a.s.l., and most of the air pollution 
sources are based on that level or even higher. The aerosol concentration normally peaks in the 
lower atmospheric layer, a few hundred meters thick, and in our case extending from 700 m 
upwards. The base station is only 100-200 m above the significant air pollution sources, so an 
impact from the greatest concentration of aerosols is not excluded in our experiment. We therefore 
expect that the performance of our method in this experiment can be representative for other sites, including places located at lower altitudes. Note that 
$AOD=0.28$ is a value as high as in most of polluted regions, and definitely much higher than 
the one typically recorded thousand of meters above ground level.

\section{Additional remarks}
\label{sec:remarks}

As pointed out above, the zenithal artificial NSB depends, among 
other parameters, on the angular emission function, the spectral power distribution, and the spatial distribution 
of the city lights. Although often imprecisely known, the values of these parameters typically change 
slowly and predictably, or are even constant from night to night in a given city. In turn, 
the highly variable columnar atmospheric optical depth has been identified as the important driving 
factor of the short-term changes in the NSB amplitude. Based on this finding we have developed a practicable 
scheme for aerosol optical depth retrieval in the lower atmosphere during a clear night, using the 
zenith sky brightness measurements acquired by two observing stations at the bottom and top of the 
atmospheric layer whose characterization is sought for. Although the specific values for $I$ and $I_{0}$ 
obtained by solving the theoretical radiative transfer models with the constraints imposed by the 
available information on the artificial sources are model-dependent, the 
ratio $I/I_{0}$ is expected to be much less sensitive to particular model choices. 

This method can be advantageously applied in clear and moonless nights in locations close to 
artificially lit nuclei, where the scattered light from artificial sources clearly outperforms 
the direct radiance from natural sources. Otherwise, the natural radiance contribution 
from the airglow, pointlike sky objects like the planets and stars, and diffuse sources 
like zodiacal light and the unresolved star background of the Milky Way and other galaxies shall be subtracted from 
the measurements. This can be done using an adequate celestial sky brightness model as, e.g. the 
ones developed by \citet[]{LeinertEtAl1998} or \citet[]{Duriscoe2013}. 

The proposed method, in combination with relatively affordable airborne platforms like drones or 
low-altitude balloons, seems to be well suited for recording continuous profiles of aerosol 
concentration at night. A large number of researchers in the academic light pollution 
research community routinely measure the night sky brightness at a wide set of 
locations distributed throughout the world. This data gathering effort is enhanced by the 
still larger number of citizen-scientists that record in a permanent way the sky brightness 
in the vicinity of their hometowns or astronomical observing sites. Overall, this paves the 
way for enhancing data sharing between the light pollution and the atmospheric research 
communities, whose fields of work are essentially intertwined. 

\section{Conclusions}
\label{sec:conclusions}

The nighttime aerosol optical depth of arbitrarily thick layers of the lower atmosphere can be 
determined by measuring the artificial sky brightness produced by city lights, avoiding that way the limitations imposed by the absence of an 
adequate reference natural light source. The method is based on the general 
properties of the solutions of the radiative transfer equation for the atmospheric propagation 
of artificial light from sources on the Earth surface, and is valid for a wide 
range of values of the aerosol optical depth. Its practical application requires relatively 
simple observational data, namely the artificial sky radiance at the zenith measured at the two
altitudes defining the limits of the layer, usually starting from a ground-based station. 
Single-channel low-cost radiance detectors can be used to that end.

A field experiment was conducted in the mountain region of High Tatra
(Slovakia). The  ground-normalized zenith brightness, along with the $AOD \approx$ 0.28 retrieved from the nearby 
AERONET station, were used to determine the aerosol content $\Delta AOD \approx$ 0.17 
of the lower atmospheric layer comprised between the two detectors (Fig.~\ref{fig:Fig3}). The mapping from the ground-normalized zenith brightness to the differential optical depth $\Delta OD = \Delta AOD + \Delta ROD$ (see Fig.~\ref{fig:Fig1}) was obtained by running a multiple scattering code \citep[]{Kocifaj2018}.
Due to its high stability, the Rayleigh-component ($ROD$) was determined 
from its theoretical formula, by assuming a molecular atmosphere scale height of 8 km. 
We have found in our experiment that $\Delta OD \approx$ 0.18 with an uncertainty 
of 0.06, related to the uncertainty of $AOD$ as described in the last paragraph in 
Sec. \ref{sec:experiment}. 
 
The aerosol optical depth is a commonly available satellite 
product, so it can advantageously be used to retrieve $\Delta AOD$ by applying the method here described. The characterization of the aerosol content of the lower atmosphere is relevant for light pollution research because it 
significantly conditions the propagation of artificial light into the nocturnal environment, and hence the range 
up to which the light from artificial sources can modify the night sky brightness. We expect this work will have a direct impact on light pollution modelling by providing 
additional information on aerosols that was largely missing for many places over the world. 
The limiting factor for the applicability of the method is the requirement that the artificial sky glow dominates the total radiance, i.e. that it be substantially larger than the natural 
background. 

\section*{Acknowledgements}

The Authors declare no conflict of interest. This work was supported by the Slovak Research and 
Development Agency under contract no: APVV-18-0014. Computational work was supported by the 
Slovak National Grant Agency VEGA (grant no. 2/0010/20). SB acknowledges support from Xunta de 
Galicia ED431B 2020/29.

\section*{Data Availability Statement}

AERONET data are publicly available on aeronet.gsfc.nasa.gov. The numerical 
results for Fig.~\ref{fig:Fig1} and the all-sky NSB were computed using the model 
available in \citet[]{Kocifaj2018}. We did not use any new data.








\bsp	
\label{lastpage}
\end{document}